\documentclass{INTERSPEECH2023}

\interspeechcameraready

\usepackage{graphicx}
\usepackage{amsmath}
\usepackage{amssymb}
\usepackage{diagbox}
\usepackage{multirow}
\usepackage{booktabs}
\usepackage{physics}
\usepackage{esvect}
\usepackage{acronym}
\usepackage[acronym]{glossaries}
\usepackage{bbding}
\usepackage{hyperref}

\usepackage{caption}
\let\OLDthebibliography\thebibliography
\renewcommand\thebibliography[1]{
  \OLDthebibliography{#1}
  \setlength{\parskip}{1.5pt}
  \setlength{\itemsep}{2pt plus 0.3ex}
}
\title{THLNet: Two-Stage Heterogeneous Lightweight Network for Monaural Speech Enhancement}
\name{Feng Dang$^{1,2,3}$ \qquad Qi Hu$^{3}$ \qquad Pengyuan Zhang$^{3}$ 
\thanks{This work was supported in part by the National Natural Science Foundation of
China (NSFC) with grant no. 11774380.}
}
  
 \address{
          $^1$Institute of Information Engineering, Chinese Academy of Sciences, Beijing, China\\
          $^2$ School of Cyber Security, University of Chinese Academy of Sciences, Beijing, China\\
          $^3$ Key Laboratory of Speech Acoustics \& Content Understanding, Institute of Acoustics, CAS, China 
          }
\email{\{dangfeng, huqi, zhangpengyuan\}@hccl.ioa.ac.cn}

\begin{document}

\ninept
\maketitle

\begin{abstract}
In this paper, we propose a two-stage heterogeneous lightweight network for monaural speech enhancement. Specifically, we design a novel two-stage framework consisting of a coarse-grained full-band mask estimation stage and a fine-grained low-frequency refinement stage. Instead of using a hand-designed real-valued filter, we use a novel learnable complex-valued rectangular bandwidth (LCRB) filter bank as an extractor of compact features. Furthermore, considering the respective characteristics of the proposed two-stage task, we used a heterogeneous structure, i.e., a U-shaped subnetwork as the backbone of CoarseNet and a single-scale subnetwork as the backbone of FineNet. We conducted experiments on the VoiceBank + DEMAND and DNS datasets to evaluate the proposed approach. The experimental results show that the proposed method outperforms the current state-of-the-art methods, while maintaining relatively small model size and low computational complexity. 
\end{abstract}

\vspace{-0.2cm}
\section{Introduction}
\vspace{-0.2cm}
\label{sec:intro}

Speech enhancement (SE) is a speech processing method that aims to improve the quality and intelligibility of noisy speech by removing noise \cite {loizou2013speech}. It is commonly used as a front-end task for automatic speech recognition, hearing aids, and telecommunications. In recent years, the application of deep neural networks (DNNs) in SE research has received increasing interest.

Many DNN-based approaches  \cite{tan2019learning, defossez20_interspeech, li2021two, dang2022dpt} have achieved impressive performance in SE tasks, but their performance improvement is accompanied by an increase in model overhead. As a result, the state-of-the-art (SOTA) models are often too large to be deployed on devices with real-world applications. Several approaches have recently been proposed to address this problem by using compact features. In RNN-noise \cite{valin2018hybrid} and PercepNet \cite{Valin2020}, bark and triangular filter banks are used to compress the spectrum, respectively. These filter banks retain the frequency domain information that is more important to human perception, effectively reducing the dimensionality of the input features and thus the complexity of the neural network model. DeepFilterNet \cite{schroter2022deepfilternet}, based on PercepNet, first enhances the spectral envelope using ERB-scaled gains and further enhances the periodic part of the preliminarily enhanced spectrum using a DeepFilter \cite{mack2019deep}. However, compact feature-based work typically uses expert hand-designed filters to derive compact real-valued features, which do not make use of phase information.

\begin{figure}
  \centering  
  \includegraphics[width=0.45\textwidth]{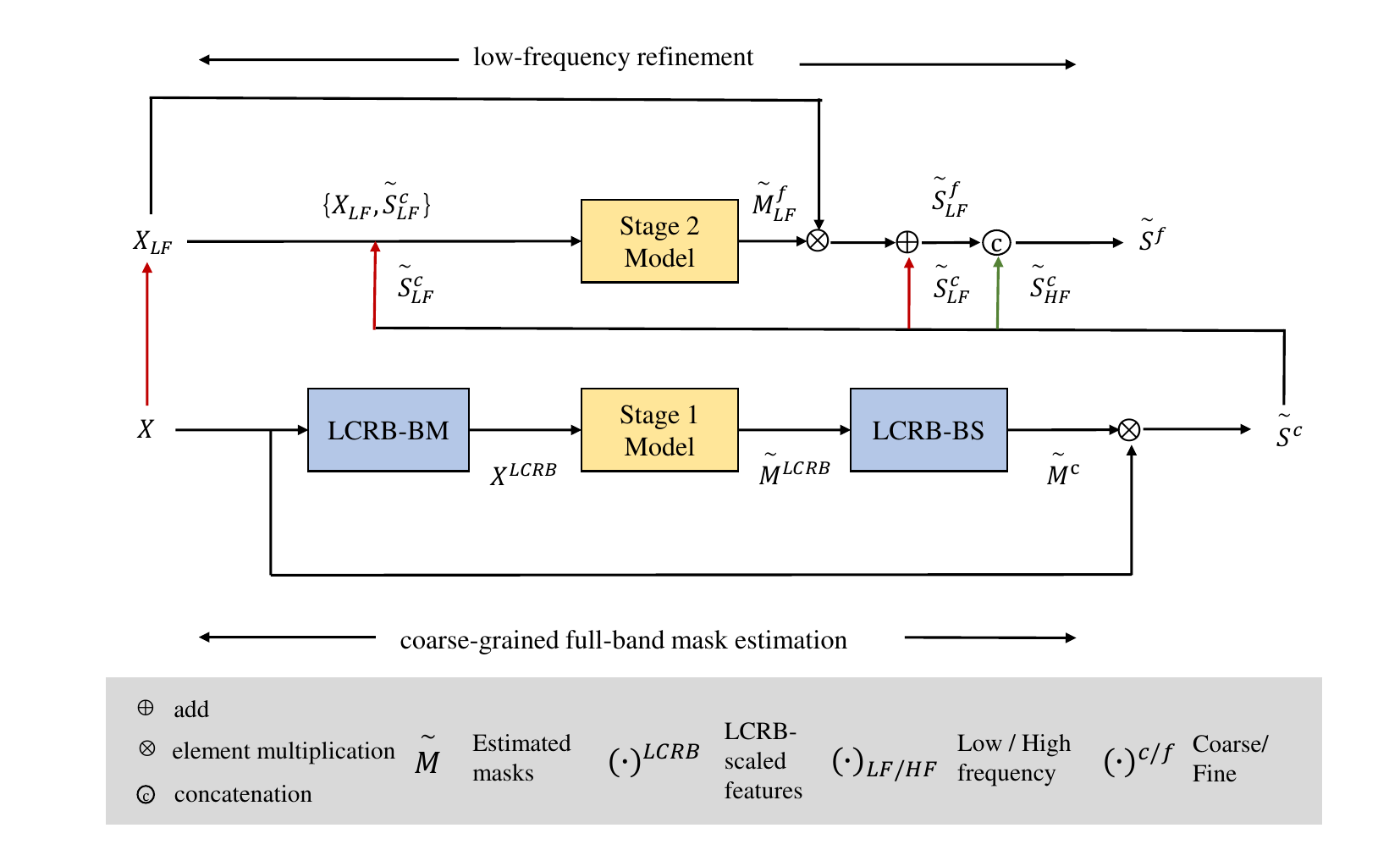}
  \vspace{-0.2cm}
  \caption{Overview of proposed system. The red and green solid arrows indicate taking the low-frequency part of the spectrum and the part other than the low frequency, respectively.}
  \label{ovl}
  \vspace{-0.7cm}
\end{figure}

Following the idea of "Divide-and-Conquer", in the multi-stage learning (MSL) approach, a difficult task is decomposed into multiple simpler sub-problems to obtain better solutions in an incremental manner and exhibit better performance than single-stage approaches in many areas, such as image inpainting \cite{hedjazi2021efficient} and image deraining \cite{li2018recurrent}. Recently, MSL has also been applied to speech front-end tasks with promising results \cite{li2021two,gao16_interspeech, hao2020masking}. Although these approaches also divide the task into easier-to-model subtasks and achieve good performance, each stage of these models essentially works on high-dimensional STFT features, resulting in a large number of parameters and computational effort.

In this context, our study makes the following contributions to the design of effective lightweight SE frameworks:
\begin{itemize}
    \item We adopt a framework that combines a two-stage task and a lightweight approach, capable of achieving comparable performance to SOTA methods with a low model overhead. Specifically, we design a novel two-stage model that includes a coarse-grained full-band mask estimation stage and a fine-grained low-frequency refinement stage \cite{dang2023first}. Instead of using a hand-designed real-valued filter, we use a novel learnable complex-valued rectangular bandwidth (LCRB) filter bank as an extractor of compact features.
    \item We adopt the idea of complementary feature processing and consider the respective characteristics of the proposed two-stage task, using a U-shaped subnetwork as the backbone of CoarseNet and a single-scale subnetwork as the backbone of FineNet.
    \item  To verify the superiority of the proposed approach, we compare our model with single-stage backbone models and other SOTA systems on two public test sets. The experimental results show that our model achieves comparable results to the single-stage backbone models with greatly reduced parameters and computational effort, and compares favorably with the SOTA models.
\end{itemize}

\vspace{-0.1cm}
\section{Proposed Algorithms}
\vspace{-0.1cm}
\label{sec:pagestyle}

The diagram of our proposed system is presented in Fig.~{\ref{ovl}}. It consists of an LCRB filter bank and two subnetworks.

In the first stage, the CoarseNet takes LCRB-scaled compact features as input and predicts an LCRB-scaled full-band complex mask $\widetilde{M}^{LCRB}$. Then, the $\widetilde{M}^{LCRB}$ is passed through the band splitting module of the LCBR filter bank to obtain the same size mask as the original spectral features (denoted as $\widetilde{M}^{c}$). In the second stage, we further refine the low-frequency part of the spectrum using the FineNet, which is fed with the low-frequency part of the complex spectrum of the original speech and the low-frequency part of the enhanced complex spectrum from the first stage (note that the low-frequency part is defined by taking the first $Q$ frequency bins). The output of the FineNet, $\widetilde{M}^{f}$, is a compensation complex mask for the low-frequency part. The final estimate $\widetilde{S}^{f}$ is obtained by summing the predicted compensation with the CoarseNet output $\widetilde{S}^{c}$. The role of FineNet is to refine the low-frequency part of the coarse signal output from CoarseNet and keep the high-frequency part of the CoarseNet signal output unchanged.

\begin{figure}
  \centering  
  \includegraphics[width=4cm]{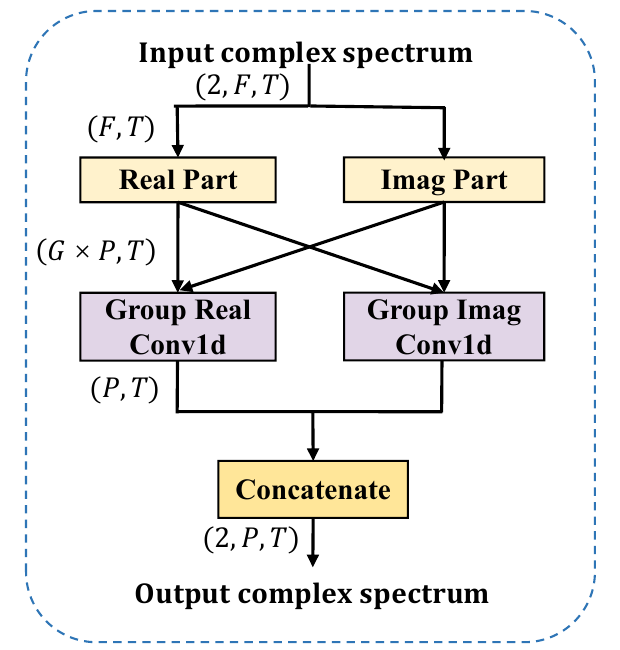}
  \vspace{-0.2cm}
  \caption{The diagram of the band merging (BM) module of the LCRB filter bank.}
  \label{LCRB}
  \vspace{-0.5cm}
\end{figure}

\vspace{-0.2cm}
\subsection{LCRB filter bank}
\label{Section21}

This paper uses a novel filter bank, denoted as LCRB,  to extract compact features. The LCRB filter bank consists of a band merging module for compressing the Fourier spectrum to obtain compact and efficient features and a band splitting module for recovering the compressed features into a Fourier spectrum. Specifically, in the band merging (BM) module, the original Fourier spectrum $X \in \mathbb{R}^{2 \times F \times T}$ is first divided into $P$ sub-bands, where each sub-band $X_{p}$ contains $G=F/P$ frequency bins, and the number of frequency bins in each sub-band is compressed to 1 by the band merging operation:
\begin{scriptsize}
\begin{equation}
	\begin{gathered}		
		X_p = X\left[:,G \ast (p - 1) + 1 : G \ast p,:\right], p=1 \cdots P, \\
        X_{p}^{’} = LCRB_{merge}(X_p), p=1 \cdots P,		
	\end{gathered}
\end{equation}
\end{scriptsize}
where $X_{p} \in \mathbb{R}^{2 \times G \times T}$ and $X_{p}^{'} \in \mathbb{R}^{2 \times 1 \times T}$. After concatenating $P$ compressed sub-band features $X_{p}^{'}$ together, we obtain a compact feature $X^{LCRB} \in \mathbb{R}^{2 \times P \times T}$ as the input to the CoarseNet encoder. The band splitting module is the inverse operation of the band merging module. The band splitting module recovers the LCRB-scaled mask $\widetilde{M}^{LCRB}$ predicted by the coarse network decoder to the Fourier spectrum mask $\widetilde{M}^{c}$. Specifically, in the band splitting (BS) module, the number of frequency bins of each sub-band mask is expanded to $G$:
\begin{scriptsize}
\begin{equation}
	\begin{gathered}		
		\widetilde{M}^{LCRB}_p = \widetilde{M}^{LCRB}\left[:,p,:\right], p=1 \cdots P, \\
       \widetilde{M}^{c}_p = LCRB_{split}(\widetilde{M}^{LCRB}_p ), p=1 \cdots P,
	\end{gathered}
\end{equation}
\end{scriptsize}
where $\widetilde{M}^{LCRB} \in \mathbb{R}^{2 \times P \times T}$, $\widetilde{M}^{LCRB}_{p} \in \mathbb{R}^{2 \times 1 \times T}$ and $\widetilde{M}^{c}_{p}  \in \mathbb{R}^{2 \times G \times T}$. After concatenating the $P$ sub-band masks $\widetilde{M}^{c}_p$, we obtain the final Fourier spectrum mask $\widetilde{M}^{c}$.    

We take advantage of the current machine learning libraries like PyTorch \cite{paszke2017automatic} to implement our module. Specifically, we can use the complex group conv1d to implement the LCRB filter bank. The band merging module of the LCRB filter bank is shown in Fig.~{\ref{LCRB}}. The band splitting module and the band merging module of the LCRB filter bank share the same structure, only the number of input and output channels of the group real (imag) conv1d need to be modified.

\begin{figure}
  \centering  
  \includegraphics[width=5.5cm]{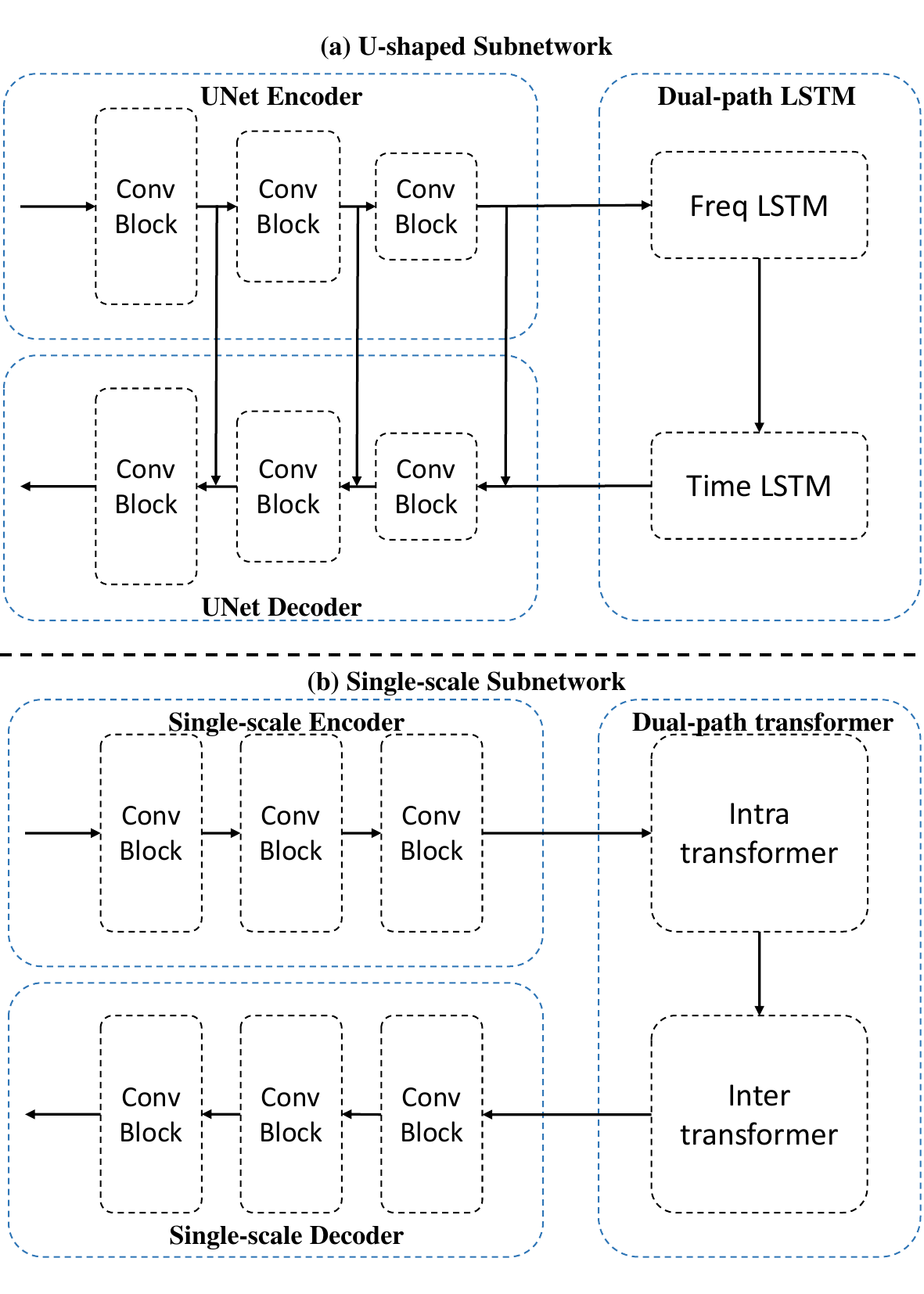}
  \vspace{-0.2cm}
  \caption{The diagram of (a) U-shaped Subnetwork. (b) Single-scale Subnetwork.}
  \label{subnetworks}
  \vspace{-0.5cm}
\end{figure}

\vspace{-0.2cm}
\subsection{Complementary feature processing}
\label{Section22}

Existing multi-stage methods for speech enhancement usually use the same structure at each stage and can be roughly classified into two categories: 1) U-shaped structures or 2) Single-scale structures. U-shaped networks \cite{li2021two,le2021dpcrn,lv2022s} first gradually map the input to a low-resolution representation and then progressively apply an inverse mapping to recover the original resolution. Although these models effectively encode multiscale information, they tend to sacrifice spatial detail due to the repeated use of downsampling operations. In contrast, methods operating on a single-scale feature pipeline are reliable in generating spectra with fine harmonic structure \cite{dang2022dpt,hao2021fullsubnet}. However, their output is less semantically robust due to either a limited receptive field or difficulty in modeling long sequences. This indicates the inherent limitations of the aforementioned architecture design choices that are capable of generating either spatially accurate or contextually reliable outputs, but not both.  Inspired by the above problem, we adopt the idea of complementary feature processing and consider the respective characteristics of the proposed two-stage task with a U-shaped subnetwork for the CoarseNet and a single-scale subnetwork for the FineNet.

Fig.~{\ref{subnetworks}}(a) and (b) shows the U-shaped subnetwork and the single-scale subnetwork we used, respectively. For the U-shaped subnetwork, we modify the DPCRN \cite{le2021dpcrn}, which consists of a U-shaped codec and multiple two-path LSTM blocks, into a lightweight version as the backbone. For the single-scale subnetwork, we modify the DPT-FSNet \cite{dang2022dpt} (DFNet for short), which consists of a single-scale dilated convolutional codec and multiple dual-path transformer blocks, into a lightweight version as the backbone.

\vspace{-0.2cm}
\subsection{Loss function}
\label{Section23}

We train the CoarseNet and FineNet jointly with the overall loss as follows:
\begin{footnotesize}
\begin{equation}
	\begin{gathered}	
     L =  L_{c} + \lambda L_{f},
	\end{gathered}
\end{equation}
\end{footnotesize}
where $\lambda$ denotes the weighted coefficient between the two losses, the loss functions of CoarseNet and FineNet, $i.e.$, $L_{c}$, $L_{f}$, are defined as:
\begin{scriptsize}
\begin{equation}
	\begin{gathered}	
     L_{c} = \alpha L_{c}^{RI} + (1-\alpha)  L_{c}^{Mag},\\
     L_{f} = \alpha L_{f}^{RI} + (1-\alpha)  L_{f}^{Mag},
	\end{gathered}
\end{equation}
\end{scriptsize}
where 
\begin{scriptsize}
\begin{equation}
	\begin{gathered}	
     {L}_{k}^{Mag}=\left | \sqrt{|\widetilde{S}_{r}^{k} |^2+ |\widetilde{S}_{i}^{k} |^2 } -  \sqrt{\left|S_{r}  \right |^2+\left |S_{i}  \right |^2    } \right |,\\
     {L}_{k}^{RI}=\left |\widetilde{S}_{r}^{k}-S_r \right | +\left |\widetilde{S}_{i}^{k}-S_i \right |,
	\end{gathered}
	\label{eq:ri+mag}
\end{equation}
\end{scriptsize}
 where $k$ takes the value of $\left\lbrace c, f \right\rbrace$ ($i.e.$, the stage label). We empirically find that $\alpha = 0.5$ and $\lambda = 1$ suffice in our evaluation.

\vspace{-0.1cm}
\section{Experiments}
\vspace{-0.1cm}

\subsection{Dataset}

We use a small-scale and a large-scale dataset to evaluate the proposed model. For the small-scale dataset, we use the VoiceBank+ DEMAND dataset \cite{valentini2016investigating}, which is widely used in SE research. This dataset contains pre-mixed noisy speech and its paired clean speech. The clean sets are selected from the VoiceBank corpus \cite{veaux2013voice}, where the training set contains 11,572 utterances from 28 speakers, and the test set contains 872 utterances from 2 speakers. For the noise set, the training set contains 40 different noise conditions with 10 types of noises (8 from DEMAND \cite{thiemann2013diverse} and 2 artificially generated) at SNRs of 0, 5, 10, and 15 dB. The test set contains 20 different noise conditions with 5 types of unseen noise from the DEMAND database at SNRs of 2.5, 7.5, 12.5, and 17.5 dB. 

For the large-scale dataset, we use the DNS dataset \cite{reddy20_interspeech}. The DNS dataset contains over 500 hours of clean clips from 2150 speakers and over 180 hours of noise clips from 150 classes. We simulate the noisy-clean pairs with dynamic mixing during the training stage. Specifically, before the start of each training epoch, $50\%$ of the clean speeches are mixed with randomly selected room impulse responses (RIR) provided by \cite{reddy2021icassp}. By mixing the clean speech ($50\%$ of them are reverberant) and noise with a random SNR in between -5 and 20 dB, we generate the speech-noise mixtures. For evaluation, we used the blind test set from the 3rd DNS Challenge, including 600 audio recordings with and without reverberation.

\vspace{-0.2cm}
\subsection{Experimental setup}

All the utterances are sampled at 16 kHz and chunked to 4 seconds for training stability. The 32 ms Hanning window is utilized, with $50\%$ overlap in adjacent frames, and 512 points FFT (Fast Fourier Transform) is used. We removed the DC component of the spectrum, resulting in 256-D (dimension) spectral features for model input. The number of bins $P$ for the input compact features of CoarseNet and $Q$ for the input low-frequency features of FineNet are set to 32 and 128, respectively.  $G(=F/P)$ is 8.

\begin{itemize}
    \item \textbf{DPCRN}: We use the configuration in \cite{le2021dpcrn} to re-implement the DPCRN, to serve as a baseline for the single-stage U-shaped network, denoted as DPCRN(base). For the modified lightweight version (denoted as DPCRN), channel number of the convolutional layers in the encoder is {64,64,64}, the kernel size and stride  are set to {(5,2), (3,2), (3,2)} and {(2,1), (2,1), (1,1)} in the frequency and time dimensions, respectively, and the hidden size of the dual-path LSTM is set to 64.

    \item \textbf{DFNet}: We use the configuration in \cite{dang2022dpt} to reimplement DFNet with some modifications to satisfy the causal settings as a baseline for the single-stage, single-scale network, denoted as DFNet(base). Causal convolution, GRU, and attention were used. In addition, in the intra transformer module (modeling sub-band temporal information), attention is removed and only the GRU is retained. For the modified lightweight version (denoted as DFNet), the difference compared to DFNet(base) is that the number of feature maps $C$ of the T-F spectrum is set to 48 and the dense connections in the codec are removed.
\end{itemize}

In the training stage, we train the proposed models for 100 epochs. We use Adam \cite{kingma2014adam} as the optimizer and a gradient clipping with a maximum L2-norm of 5 to avoid gradient explosion. The initial learning rate (LR) was set to 0.0004 and then decayed by a factor of 0.98 every two epochs.  Our aim is to design lightweight real-time online models where all operations in the models (except those in the inter transformer) are causal and the overall latency of the algorithm is 48ms. The core codes are available\footnote{https://github.com/dangf15/THLNet}.

\begin{table}[t]
\caption{Ablation analysis results on the VoiceBank+DEMAND dataset}
\label{tab:2}
\centering
\resizebox{0.45\textwidth}{13mm}{
\begin{tabular}{c c c c c c}
\toprule  
 
Method  &  Params(M) / MACS(G)  & WB-PESQ & STOI & SI-SDR \\
\midrule
DFNet(base)   & 0.77 / 13.60 & 3.06 & 94.95 & 19.06   \\
DPCRN(base)  & 1.02 / 4.67 & 3.01 & 94.78 & 18.65   \\
\midrule
2DFNet   & 0.54  / 2.97 & 2.87 & 94.26 & 18.67   \\
2DPCRN   & 0.61  / 1.13 & 2.92 & 94.55 & 18.71   \\
\midrule
DFNet+DPCRN  & 0.58  / 1.47 & 2.95 & 94.48 & 18.90   \\
DPCRN+DFNet  & 0.58  / 2.63 & 3.04 & 94.67 & 19.04   \\
\bottomrule
\end{tabular}}
\label{tbl:ablation-study1}
\vspace{-.2cm}
\end{table}

\renewcommand\arraystretch{1}
\begin{table}[t]
\caption{Ablation analysis results on the DNS dataset}
\scriptsize
\label{tab:3}
\centering

\begin{tabular}{c c c c c}
\toprule  
 
Method & Type & nSIG  & nBAK & nOVL \\
\midrule
Noisy & - & 3.87  & 3.05 & 3.11 \\
DFNet(base) & S  & 3.95  & 4.33 & 3.64   \\
DPCRN(base) & U & 3.92  & 4.24 & 3.53 \\
\midrule
2DFNet  & S+S & 3.90  & 4.23 & 3.51  \\
2DPCRN  & U+U & 3.91  & 4.28 & 3.57  \\
\midrule
DFNet+DPCRN & U+S & 3.93  & 4.29 & 3.57  \\
DPCRN+DFNet & S+U & 3.94  & 4.32 & 3.61  \\
\bottomrule
\end{tabular}
\label{tbl:ablation-study2}
\vspace{-.4cm}
\end{table}

\vspace{-0.2cm}
\subsection{Evaluation metrics}


For the VoiceBank+DEMAND dataset, we use wide-band PESQ (denoted as WB-PESQ), STOI, CSIG, CBAK, and COVL  as evaluation metrics. For the DNS dataset, we used the subjective results predicted by DNSMOS P. 835 \cite{reddy2022dnsmos}. The three predicted scores are denoted as nOVL,nSIG,and nBAK, which predict the overall subjective score, the quality score of speech distortion, and the quality score of noise removal, respectively.

\begin{figure}
  \centering  
  \includegraphics[width=7.5cm]{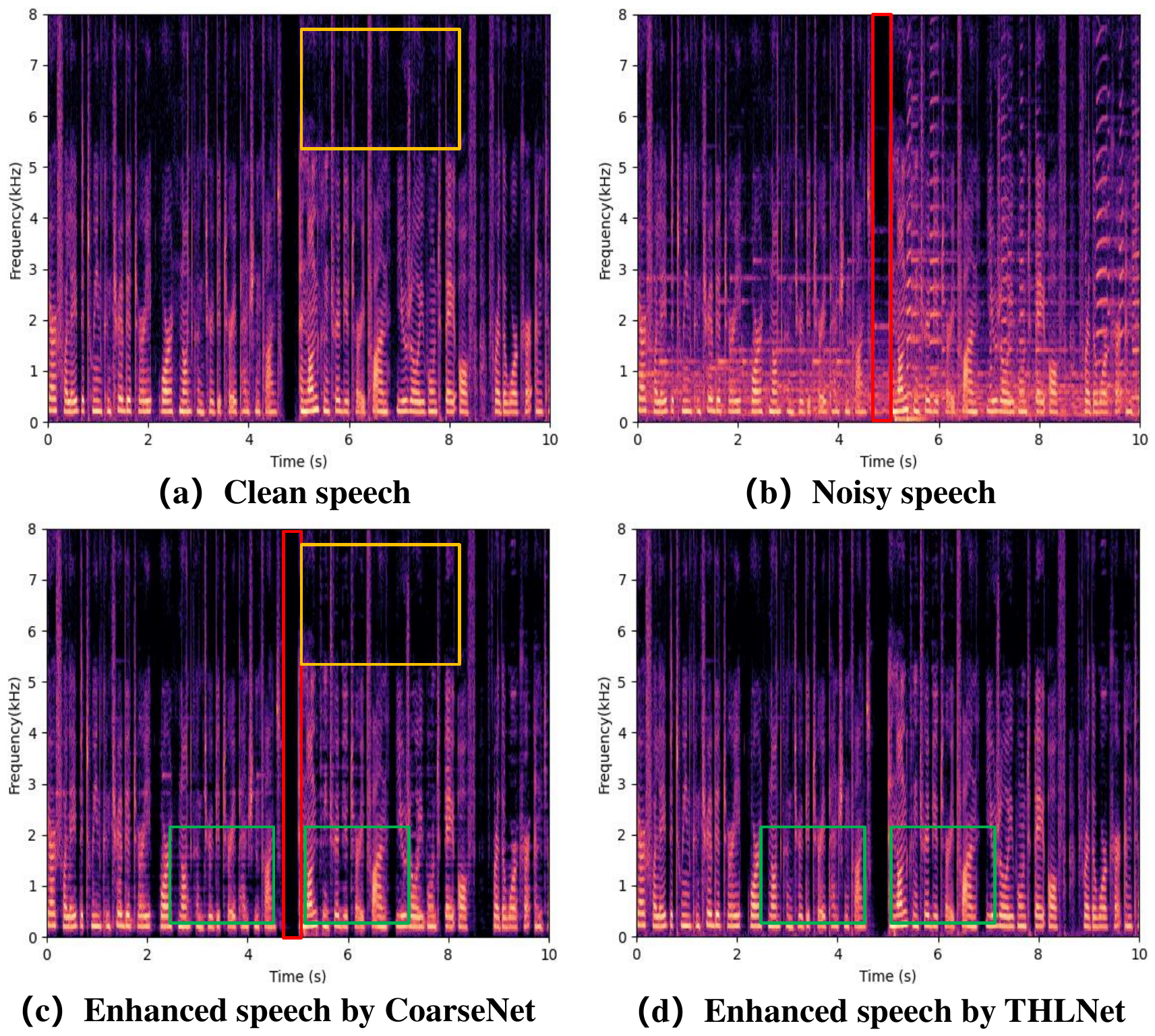}
  \vspace{-0.1cm}
  \caption{An illustration of the enhanced results using our proposed models.  The speech spectrogram of \textbf{(a)} the clean utterance. \textbf{(b)} the noisy utterance. \textbf{(c)}  the enhanced utterance by CoarseNet. \textbf{(d)} the enhanced utterance by THLNet.}
  \label{specs}
  \vspace{-0.4cm}
\end{figure}

\vspace{-0.1cm}
\section{Evaluation Results}
\vspace{-0.1cm}

\renewcommand\arraystretch{1.0}
\begin{table*}[t]
\caption{Comparison with other state-of-the-art causal systems on the VoiceBank+DEMAND dataset.}
\scriptsize
\label{tab:1}
\centering

\begin{tabular}{c c c c c c c c c}
\toprule  
 
Method  & Params(M) & MACS(G) & WB-PESQ  & STOI & CSIG & CBAK & COVL    \\
\midrule
Noisy  & - & - & 1.97  & 0.92 & 3.34 & 2.44 & 2.63   \\
\midrule
PercepNet~{\cite{Valin2020}}  & 8.00 & {0.80} & 2.73  & - & - & - & -  \\
DCCRN~{\cite{Hu2020}}   & 3.70 & 14.36  & 2.54 & 0.94 & 3.74 & 3.13 & 2.75 \\
DEMUCS~{\cite{defossez20_interspeech}}  & 18.9 & -  & 2.93  & 0.95 & 4.22 & 3.25 & 3.52  \\
CTS-Net~{\cite{li2021two}}   & 4.35 & 5.57 & 2.92  & - & 4.25 & 3.46 & 3.59  \\
S-DCCRN~{\cite{lv2022s}}  & 2.34 & -  & 2.84  & 0.94 & 4.03 & 3.43 & 2.97  \\
FullSubNet+~{\cite{chen2022fullsubnet+}}  & 8.67 & 30.06  & 2.88  & 0.94 & 3.86 & 3.42 & 3.57  \\
\midrule
CoarseNet(Pro.)  & \textbf{0.31} & \textbf{0.22}  & 2.61 & 0.93 & 3.89 & 3.25 & 3.24  \\
THLNet(Pro.)   & {0.58}  & {2.63} & \textbf{3.04}  & \textbf{0.95} & \textbf{4.27} & \textbf{3.59} & \textbf{3.66} \\
\bottomrule
\end{tabular}
\label{tbl:sota}
\vspace{-.4cm}
\end{table*}

\subsection{Ablation analysis}

Table~{\ref{tbl:ablation-study1}} and Table~{\ref{tbl:ablation-study2}} show the results of the ablation experiments on the VoiceBank+DEMAND dataset and the DNS dataset, respectively. We classify these models into three types: 1) single-stage models, i.e., DFNet(base) and DPCRN(base); 2) two-stage homogeneous models, i.e., 2DFNet and 2DPCRN; and 3) two-stage heterogeneous models, i.e., DFNet+DPCRN and DPCRN+DFNet. The sub-models of each stage are further divided into two categories, i.e., single-scale subnetworks and U-shaped subnetworks, abbreviated as S and U, as shown in Table~{\ref{tbl:ablation-study2}},. From Table~{\ref{tbl:ablation-study1}} and Table~{\ref{tbl:ablation-study2}}, the following observations can be obtained.

\begin{itemize}
    \item By comparing 2DFNet/2DPCRN and DFNet(base)/ DPCRN (base), it can be seen that decomposing the task into two stages and using two lightweight subnetworks with the same structure results in acceptable performance degradation over a single-stage model using the same topology, but greatly reduces the number of multiply-accumulate operations (MACS).
    \item By comparing DFNet+DPCRN/DPCRN+DFNet and 2DFNet/ 2DPCRN, it can be seen that there is an observable improvement in the performance of the two stages using heterogeneous subnetworks over the two stages using homogeneous subnetworks.
    \item By comparing DFNet+DPCRN/DPCRN+DFNet and DFNet (base)/DPCRN(base), it can be seen that when the two-stage model uses heterogeneous subnetworks, comparable performance to the best single-stage baseline model (i.e., DFNet(base)) can be achieved, but greatly reduces the number of multiply-accumulate operations.
\end{itemize}

The best performing two-stage model is DPCRN+DFNet, probably because the first stage models coarse-grained full-band information, i.e., it mainly needs to model the contextual information of the entire spectrum, which makes DPCRN more suitable due to the multiscale nature of the UNet structure. The second stage, on the other hand, mainly plays the role of fine-grained low-frequency spectrum recovery, and the use of a single-scale subnetwork DFNet that does not involve downsampling (i.e., no distortion of harmonic information) is more appropriate. Compared with DFNet(base), DPCRN+DFNet achieves comparable scores for objective metrics, i.e., WB-PESQ, STOI, SI-SDR, and subjective metrics, i.e., nSIG,nBAK,nOVL, while the number of parameters (Params) and MACS is 75.3$\%$ and 19.3$\%$ of those of DFNet(base), respectively.

\vspace{-0.2cm}
\subsection{Comparison with other SOTA causal methods}

Table~{\ref{tbl:sota}} shows the comparison of our proposed model (the best performing sub-network combination \textit{\textbf{DPCRN+DFNet}}) with other SOTA causal systems on the VoiceBank+DEMAND dataset. As can be seen from the table, among these advanced causal systems, THLNet has the highest scores in all evaluation metrics with relatively small parameters and MACS.

\vspace{-0.1cm}
\section{Discussions}
\vspace{-0.1cm}

From Fig.~{\ref{specs}}, we can see that the CoarseNet can effectively suppress the noise components and recover the main geometric structure of the high-frequency part of the spectrum. For example, the CoarseNet achieves good performances under background noise conditions, as shown in the red marker area in Fig.~{\ref{specs}} (b) and (c), and restores the main geometric structure of the high-frequency part of the spectrum as shown in the yellow marker area of Fig.~{\ref{specs}} (a) and (c). However, the CoarseNet is poor at handling the details of the harmonic structure in the low-frequency part of the spectrum. Fortunately, the low-frequency harmonic structure can be effectively recovered by the FineNet, as shown in the green marker area of Fig.~{\ref{specs}} (c) and (d). A speech of fairly good quality is obtained by the two-stage processing with the CoarseNet and the FineNet.

\vspace{-0.1cm}
\section{Conclusions}
\vspace{-0.1cm}

We propose THLNet, a two-stage heterogeneous lightweight network for monaural speech enhancement. The network consists of an LCRB filter bank and two mask-based subnetworks, CoarseNet and FineNet, which are responsible for the coarse-grained full-band mask estimation and the fine-grained low-frequency refinement, respectively. We use a novel LCRB filter bank as an extractor of compact features. Further, considering the different characteristics of the proposed two-stage task, we use a heterogeneous structure, i.e., a U-shaped subnetwork (DPCRN) as the backbone of CoarseNet and a single-scale subnetwork (DFNet) as the backbone of FineNet, which further improves the performance of the proposed algorithm. Experimental results on the VoiceBank + DEMAND dataset show that the proposed approach outperforms the current SOTA,  while maintaining a relatively small model size and low computational complexity.

\bibliographystyle{IEEEtran}
\bibliography{ref}

\end{document}